\documentclass[12pt]{article}

\pdfoutput=1

\usepackage{jheppub}
\usepackage{graphicx,epsfig,amsmath,amssymb}
\usepackage{subcaption}
\usepackage{multirow}
\usepackage{slashed}
\usepackage{cleveref}
\usepackage{enumitem}
\usepackage{calc}

\def\be{\begin{equation}}
\def\ee{\end{equation}}
\def\bea{\begin{eqnarray}}
\def\eea{\end{eqnarray}}

\newcommand{\mhh}{m_{hh}}

\newcommand{\ct}{c_t}
\newcommand{\ctt}{c_{tt}}
\newcommand{\chhh}{c_{hhh}}
\newcommand{\cg}{c_{ggh}}
\newcommand{\cgg}{c_{gghh}}

\title{SMEFT predictions for $gg\to hh$ at full NLO QCD and truncation uncertainties}
\author[a]{Gudrun Heinrich,}
\author[a]{Jannis Lang,}
\author[b]{Ludovic Scyboz}

\affiliation[a]{Institute for Theoretical Physics, Karlsruhe Institute of Technology (KIT), 76131 Karlsruhe, Germany}
\affiliation[b]{Rudolf Peierls Centre for Theoretical Physics, Parks Road, Oxford OX1 3PU, UK}

\emailAdd{gudrun.heinrich@kit.edu}
\emailAdd{jannis.lang@kit.edu}
\emailAdd{ludovic.scyboz@physics.ox.ac.uk}

\preprint{{\small KA-TP-14-2022\\
    \hphantom{.}\hfill OUTP-22-05P\\
    \hphantom{.}\hfill  P3H-22-045}}

\abstract{
 We present a calculation of the NLO QCD corrections for Higgs-boson
 pair production in gluon fusion including effects of anomalous
 couplings within Standard Model Effective Field Theory (SMEFT).
 We study effects of different truncation options of the EFT expansion in $1/\Lambda$
and of double operator insertions, both at total cross-section level as well
as for the distribution of the invariant mass of the Higgs-boson pair,
at $\sqrt{s}=13$\,TeV.
 The NLO corrections are implemented in the generator
 \texttt{ggHH\_SMEFT} in the \texttt{Powheg-Box-V2} framework.
}


\begin{document}

\maketitle

\section{Introduction}

The importance of Higgs-boson pair production as a process allowing us to shed
more light on the Higgs potential  is undisputed.  Deviations from the Standard
Model (SM) form, manifesting themselves in anomalous Higgs-boson self-couplings,
would be a clear sign of new physics. If the Higgs-boson
trilinear coupling was found to be different from the SM value, small deviations in other Higgs couplings could also be expected. In order to expose such slight deviations, it is crucial to control the uncertainties of the theory
predictions, including the description of anomalous couplings within an
effective field theory (EFT) framework.
The theory uncertainties have various sources, the dominant ones in the SM
currently being uncertainties related to the treatment of the top-quark mass in
different renormalisation schemes.
NLO QCD corrections including the full top-quark mass dependence are
available~\cite{Borowka:2016ehy,Borowka:2016ypz,Baglio:2018lrj,Baglio:2020ini}
and have been included in calculations where higher orders have been performed
in the heavy-top limit~\cite{Grazzini:2018bsd,Chen:2019lzz,Chen:2019fhs}, thus
reducing the scale uncertainties and the uncertainties due to missing top-quark
mass effects, while the top-mass scheme uncertainties remain an
issue~\cite{Baglio:2020wgt}.

Going beyond the SM description of the process $gg\to hh$, effects of anomalous
couplings have been studied at NLO in the Born-improved heavy-top limit (HTL)
for both CP-conserving~\cite{Grober:2015cwa} as well as
CP-violating~\cite{Grober:2017gut} operators. NLO corrections with full $m_t$-dependence have been incorporated within a non-linear EFT parametrisation (also called
Higgs Effective Field Theory, HEFT) in Ref.~\cite{Buchalla:2018yce}. In
Ref.~\cite{deFlorian:2017qfk} the combination of NNLO corrections in the HTL
has been performed within the HEFT framework. Finally, in
Ref.~\cite{deFlorian:2021azd}, the full NLO corrections of
Ref.~\cite{Buchalla:2018yce} have been combined with the NNLO corrections of
Ref.~\cite{deFlorian:2017qfk} to provide approximate NNLO predictions, dubbed
NNLO$^\prime$, which include the full top-quark mass dependence up to NLO and
higher order corrections up to NNLO in the HTL, combined with operators related
to the five most relevant anomalous couplings for the process $gg\to hh$.

By including the EFT parametrisation of new physics effects into the predictions for
Higgs-boson pair production, new uncertainties arise, related to the truncation
of the EFT expansion, which relies on an assessment of the relevance of operator contributions to the Lagrangian in a certain well-defined counting scheme.
Furthermore, at amplitude squared level, there are several possibilities to
truncate the expansion in the canonical dimension, related to the inclusion of
squared dimension six terms and double operator insertions.
The discussion of truncation uncertainties recently gained
considerable attention in the literature~\cite{Biekotter:2016ecg,Brivio:2022pyi,Dawson:2021xei,Lang:2021hnd,Ethier:2021bye,Trott:2021vqa,Martin:2021cvs,Battaglia:2021nys,Aoude:2022imd},
showing that the uncertainties have to be assessed on a case-by-case basis.

In the following we will present results within SMEFT for Higgs-boson pair
production in gluon fusion, including full NLO QCD corrections as well
as options to include linear or quadratic terms in the EFT expansion and double operator insertions.
This allows us to investigate various scenarios of truncation of the
EFT expansion and to assess the related uncertainties compared to the size of the scale uncertainties at NLO QCD.
We will also point out differences between a SMEFT and a HEFT description in this context.
 In \cref{sec:EFTframework}, we recap the
definition of the SMEFT and HEFT effective field-theory frameworks, specifically
of those operators that are of interest for Higgs-pair production, and we
outline different ways of defining the truncation at the level of the squared
amplitude. In~\cref{sec:usage} we give a brief summary of the implementation of
the SMEFT predictions at full NLO QCD into the {\tt Powheg-Box-V2}
~\cite{Nason:2004rx,Frixione:2007vw,Alioli:2010xd} event 
generator. We discuss our results in~\cref{sec:results}, and conclude
in~\cref{sec:conclusions}.

\section{Description of the EFT framework and the calculation}
\label{sec:EFTframework}

\subsection{EFT descriptions of Higgs-boson pair production}
\label{sec:HeftSmeft}

In this section, we introduce our conventions and contrast the SMEFT and HEFT
descriptions at Lagrangian level.

In Standard Model Effective Field Theory
(SMEFT)~\cite{Buchmuller:1985jz,Grzadkowski:2010es,Brivio:2017vri}, a low energy
description of unknown interactions at a new physics scale $\Lambda$ is
constructed as an expansion in inverse powers of $\Lambda$, with operators
$\mathcal{O}_i$ of canonical dimension larger than four and corresponding Wilson
coefficients $C_i$,
\begin{equation}
\mathcal{L}_\text{SMEFT} = \mathcal{L}_\text{SM} + \sum_{i}\frac{C_i^{(6)}}{\Lambda^2}\mathcal{O}_i^{\rm{dim6}} +{\cal O}(\frac{1}{\Lambda^3})\; .
\label{eq:Ldim6}
\end{equation}
In SMEFT it is assumed that the physical Higgs boson is part of a doublet
transforming linearly under $SU(2)_L\times U(1)$.  The SMEFT Lagrangian is
usually given in the so-called Warsaw basis~\cite{Grzadkowski:2010es}, where the
terms relevant to the process $gg\to hh$ read
\begin{equation}
\begin{split}
\Delta\mathcal{L}_{\text{Warsaw}}&=\frac{C_{H,\Box}}{\Lambda^2} (\phi^{\dagger} \phi)\Box (\phi^{\dagger } \phi)+ \frac{C_{H D}}{\Lambda^2}(\phi^{\dagger} D_{\mu}\phi)^*(\phi^{\dagger}D^{\mu}\phi)+ \frac{C_H}{\Lambda^2} (\phi^{\dagger}\phi)^3 \\ &+\left( \frac{C_{uH}}{\Lambda^2} \phi^{\dagger}{\phi}\bar{q}_L\phi^c t_R + h.c.\right)+\frac{C_{H G}}{\Lambda^2} \phi^{\dagger} \phi G_{\mu\nu}^a G^{\mu\nu,a}\;.  \label{eq:warsaw}
\end{split}
\end{equation}
The dipole operator $\bar{{\cal O}}_{uG}$ is not included here because it can be
shown that it carries an extra loop suppression factor $1/(4\pi)^2$ relative to
the other contributions if weak coupling to the heavy sector is
assumed~\cite{Buchalla:2022vjp,Buchalla:2018yce,Arzt:1994gp}. In the case of a UV completion where the
coupling to the heavy sector is strong, SMEFT would not be the appropriate
description of the full theory at low energies anyway.

\vspace*{3mm}

Higgs Effective Field Theory
(HEFT)~\cite{Feruglio:1992wf,Burgess:1999ha,Grinstein:2007iv,Contino:2010mh,Alonso:2012px,Buchalla:2013rka}
instead is based on an expansion in terms of loop orders, which also can be
formulated in terms of chiral dimension ($d_\chi$)
counting~\cite{Weinberg:1978kz,Buchalla:2013eza,Krause:2016uhw}. The expansion parameter is given
by $f^2/\Lambda^2\simeq \frac{1}{16\pi^2}$, where $f$ is a typical energy scale
at which the EFT expansion is valid (such as the pion decay constant in chiral
perturbation theory),
\begin{align}
  {\cal L}_{d_\chi}={\cal L}_{(d_\chi=2)}+\sum_{L=1}^\infty\sum_i\left(\frac{1}{16\pi^2}\right)^L c_i^{(L)} O^{(L)}_i\;.
  \label{eq:loop_expansion}
  \end{align}

The HEFT Lagrangian relevant to Higgs-boson pair production in gluon fusion can
be parametrised by five a priori independent anomalous couplings as
follows~\cite{Buchalla:2018yce}
\begin{align}
\Delta{\cal L}_{\text{HEFT}}=
-m_t\left(c_t\frac{h}{v}+c_{tt}\frac{h^2}{v^2}\right)\,\bar{t}\,t -
c_{hhh} \frac{m_h^2}{2v} h^3+\frac{\alpha_s}{8\pi} \left( c_{ggh} \frac{h}{v}+
c_{gghh}\frac{h^2}{v^2}  \right)\, G^a_{\mu \nu} G^{a,\mu \nu}\;.
\label{eq:ewchl}
\end{align}

Expanding the Higgs doublet in eq.~\eqref{eq:warsaw} around its vacuum
expectation value and applying a field redefinition to the physical Higgs boson
\begin{align}
h\to h+v^2\frac{C_{H,\textrm{kin}}}{\Lambda^2}\left(h+\frac{h^2}{v}+\frac{h^3}{3v^2}\right)\;,\label{eq:field_redefinition}
\end{align}
with $$C_{H,\textrm{kin}}:=C_{H,\Box}-\frac{1}{4}\,C_{HD}\;,$$
the Higgs kinetic term acquires its canonical form (up to ${\cal O}\left(\Lambda^{-4}\right)$ terms).
After that, the couplings can be related through a comparison of the coefficients of the
corresponding terms in the Lagrangian, which leads to the expressions given
in Table~\ref{tab:translation}. Note that in the Warsaw basis $C_{HG}$
implicitly contains a factor of $\alpha_s$ relative to $\cg$ and $\cgg$ and
therefore the translation becomes scale-dependent even if no (electroweak) RGE
running of the Wilson coefficients is included.

\begin{table}[htb]
\begin{center}
\begin{tabular}{ |c |c| }
\hline
HEFT&Warsaw\\
\hline
$c_{hhh}$ & $1-2\frac{v^2}{\Lambda^2}\frac{v^2}{m_h^2}\,C_H+3\frac{v^2}{\Lambda^2}\,C_{H,\textrm{kin}}$ \\
\hline
$c_t$ &  $1+\frac{v^2}{\Lambda^2}\,C_{H,\textrm{kin}} - \frac{v^2}{\Lambda^2} \frac{v}{\sqrt{2} m_t}\,C_{uH}$\\
\hline
$ c_{tt} $ & $-\frac{v^2}{\Lambda^2} \frac{3 v}{2\sqrt{2} m_t}\,C_{uH} + \frac{v^2}{\Lambda^2}\,C_{H,\textrm{kin}}$\\
\hline
$c_{ggh}$ &  $\frac{v^2}{\Lambda^2} \frac{8\pi }{\alpha_s} \,C_{HG}$ \\
\hline
$c_{gghh}$ &  $\frac{v^2}{\Lambda^2}\frac{4\pi}{\alpha_s} \,C_{HG}$ \\
\hline
\end{tabular}
\end{center}
\caption{Translation at Lagrangian level between different operator basis choices.\label{tab:translation}}
\end{table}

The translation given in Table~\ref{tab:translation} suggests that
there is no explicit dependence on the scale $\Lambda$ in the HEFT Wilson
coefficients. As mentioned above, the effective HEFT Lagrangian is expanded in
powers of $f^2/\Lambda^2 \simeq 1/(4\pi)^2$, with $\Lambda \simeq 4 \pi f$ the
scale of new physics
and $f$ a reference scale for energies where the EFT expansion is valid; for the case of strongly coupled UV completions $f$ corresponds to the scale of dynamical symmetry breaking.
In section~\ref{sec:results}, we will still use the translation for a specific value of the scale
$\Lambda$ to compare SMEFT and HEFT results.

However, it should be pointed out that a translation between the coefficients at
Lagrangian level must be applied with care.  The EFT parametrisations have a
validity range limited by unitarity constraints and the assumption that
$C_i/\Lambda^2$ in SMEFT is a small quantity.  Furthermore, due to different
assumptions about the transformation of the Higgs field under the EW symmetry
transformations, there are relations between certain coefficients in SMEFT, which are
not present in HEFT. Therefore a naive translation from HEFT (which is, in this regard, the more
general theory) to SMEFT can lead out of the validity range of SMEFT for certain
points in the coupling parameter space, even though they are perfectly valid points
in HEFT.

\subsection{Operator insertions at amplitude-squared level}
\label{sec:truncation}

An EFT description is based on an expansion in a parameter encoding the scale
hierarchies which underlie the EFT description.  Therefore an uncertainty arises
due to the truncation of the EFT expansion.  In particular, there is the
question whether to truncate the SMEFT expansion at amplitude-squared level
strictly beyond dimension-6, or to include squared dimension-6 operators, which are
of order $1/\Lambda^4$ and therefore formally suppressed at the same order as dimension-8
operators. Furthermore,  double operator insertions into a single diagram are
usually neglected as they form a subset of operators at order $1/\Lambda^4$
which is not uniquely defined (for example, they can be related to a different
set of operators through the equations of motion). Such issues have been
discussed recently in
Refs.~\cite{Brivio:2022pyi,Dawson:2021xei,Battaglia:2021nys,Trott:2021vqa,Martin:2021cvs}.

In the next section we present a Monte Carlo program, which includes the full
$m_t$-dependent NLO QCD corrections, and allows the systematic study of
truncation effects for the case of Higgs-boson pair production in gluon fusion.
In order to construct the different truncation options we divide the amplitude
into three parts: the pure SM contribution ($\text{SM}$), single dimension-6
operator insertions ($\rm{dim6}$) and double dimension-6 operator insertions
($\rm{dim6}^2$):
\begin{align}
{\cal M}=
&\ 
\vcenter{\hbox{\includegraphics[page=1,scale=0.9]{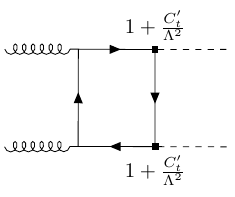}}}
+\vcenter{\hbox{\includegraphics[page=2,scale=0.9]{figures/gghh_diagrams}}}
+\vcenter{\hbox{\includegraphics[page=3,scale=0.9]{figures/gghh_diagrams}}}
\nonumber\\
&\ +\vcenter{\hbox{\includegraphics[page=4,scale=0.9]{figures/gghh_diagrams}}}
+\vcenter{\hbox{\includegraphics[page=5,scale=0.9]{figures/gghh_diagrams}}}
+\dots
\nonumber\\
=&\ 
{\cal M}_\text{SM} + {\cal M}_{\rm{dim6}} + {\cal M}_{\rm{dim6}^2}\;,\label{eq:amplitude_expansion}
\end{align}
where $C^\prime$ denotes the corresponding coupling combination listed in
Table~\ref{tab:translation}.
We consider four
possibilities to choose which parts of $|{\cal M}|^2$ from
eq.~\eqref{eq:amplitude_expansion} may enter in the squared amplitude forming the cross section:
\begin{align}
\sigma \simeq \left\{\begin{aligned}
&\ \sigma_\text{SM} + \sigma_{\text{SM}\times \rm{dim6}} &\textrm{(a)}
\\
&\ \sigma_{\left(\text{SM}+\rm{dim6}\right)\times \left(\text{SM}+\rm{dim6}\right)}  &\textrm{(b)}
\\
&\ \sigma_{\left(\text{SM}+\rm{dim6}\right)\times \left(\text{SM}+\rm{dim6}\right)}  + \sigma_{\text{SM}\times \rm{dim6}^2} &\textrm{(c)}
\\
&\ \sigma_{\left(\text{SM}+\rm{dim6}+\rm{dim6}^2\right)\times \left(\text{SM}+\rm{dim6}+\rm{dim6}^2\right)} &\textrm{(d)}
\end{aligned}\right.\label{eq:truncation}
\end{align}
Case (a) denotes the first order of an expansion of $\sigma\sim |{\cal M}|^2$
in $\Lambda^{-2}$, (b) is the first order of an expansion of ${\cal M}$ in
$\Lambda^{-2}$.  Case (c) includes all terms of ${\cal
O}\left(\Lambda^{-4}\right)$ coming from single and double dimension-6 operator
insertions, however it lacks the contribution at the same order from
dimension-8 operators and ${\cal O}\left(\Lambda^{-4}\right)$ terms following
from the field redefinition of eq.~\eqref{eq:field_redefinition}.  Case (d)
would correspond to HEFT upon using the translation of the parameters as in
Table~\ref{tab:translation}, except for differences due to the running in
$\alpha_s$, because no linearisation whatsoever in $1/\Lambda$ is present
there.

\section{Implementation and usage of the code within the {\tt Powheg-Box}}
\label{sec:usage}

\subsection{Implementation of the NLO QCD corrections}

Our implementation is similar to the NLO HEFT code {\tt Powheg-Box-V2/ggHH},
which was presented in Refs.~\cite{Heinrich:2019bkc,Heinrich:2020ckp} and is
publicly available.  For the virtual two-loop corrections, the NLO QCD
corrections in the SM as calculated in
Refs.~\cite{Borowka:2016ehy,Borowka:2016ypz} have been used and extended to the
SMEFT framework. The operator insertions have been included in a modular way,
such that the different options described in the previous section can be
calculated.


For the real emission part, the one-loop $2\to 3$ matrix elements have been
produced with {\tt GoSam}~\cite{Cullen:2011ac,Cullen:2014yla}, based on a model
which we generated in UFO format~\cite{Degrande:2011ua}, specifying the
anomalous couplings such that {\tt GoSam} is able to calculate the different
contributions according to the chosen truncation option.  The existing
interface~\cite{Luisoni:2013cuh} to {\tt
Powheg}~\cite{Nason:2004rx,Frixione:2007vw,Alioli:2010xd} has been modified to hand over
event parameters to {\tt GoSam} in such a way that the factor $\alpha_s$, which
is included in the definition of the Higgs-gluon couplings in SMEFT, is
evaluated at the correct energy scale. The renormalisation of the top-quark
mass is performed in the on-shell scheme.

The code is built such that it splits the amplitude evaluation according to
eq.~\eqref{eq:amplitude_expansion}, and the squared amplitude is calculated with
truncation option (a), (b), (c) or (d) as defined by the user in the input card.
 
\subsection{Usage within the {\tt Powheg-Box-V2}}

Usage of the program {\tt ggHH\_SMEFT} is similar to the HEFT public code {\tt
ggHH}~\cite{Heinrich:2020ckp}, and both are provided within the {\tt
POWHEG-BOX-V2}~\cite{Alioli:2010xd} under {\tt User-Processes-V2}.  The user defines
the EFT parameters directly in the input card, of which we provide an
example in {\tt testrun/powheg.input-save}. The EFT mass scale is set by:
\begin{description}[leftmargin=!]
  \item[\qquad{\tt Lambda=1.0}\,:] { the input value of the SMEFT heavy mass
  scale $\Lambda$ (in TeV).}
\end{description}
The value of the SMEFT coefficients in the Warsaw basis is set by the following
keywords:
\begin{description}[leftmargin=!,labelwidth=\widthof{\qquad{\tt Lambda=1.0}\,: }]
  \itemsep0em 
  \item[\qquad{\tt CHbox}\,:] { the Higgs kinetic term coefficient $C_{H,\Box}$,}
  \item[\qquad{\tt CHD}\,:] { the Higgs kinetic term coefficient $C_{HD}$, }
  \item[\qquad{\tt CH}\,:] { the Higgs trilinear coupling term $C_H$,}
  \item[\qquad{\tt CuH}\,:] { the Higgs Yukawa coupling to up-type quarks term $C_{uH}$,}
  \item[\qquad{\tt CHG}\,:] { the effective coupling of gluons to Higgs bosons $C_{HG}$.}
\end{description}
The input keyword \texttt{usesmeft} can be set to \texttt{0} or \texttt{1}. When
{\tt usesmeft=1}, the above parameters for {\tt CHbox, CHD, CH, CuH, CHG} are
taken from the input card and translated internally to be used in the
computation of the amplitude. If {\tt usesmeft=0}, the values of the parameters
{\tt chhh, ct, ctt, cggh, cgghh} are used instead. If {\tt usesmeft=1}
is set, the values for the parameters {\tt chhh, ct, ctt, cggh, cgghh}
will be ignored.

Finally, the different SMEFT truncation options can be selected via the keyword
{\tt multiple-insertion}, where the options (a)--(d) in
eq.~\eqref{eq:truncation} correspond to the values 0--3 of this flag. Apart from
the above, the usage of the code is as described in~\cite{Heinrich:2020ckp} and
in the {\tt Docs} folder of the code. In particular, we remind the reader that
the 2-loop grids for the virtual contributions have been generated with a {\it
fixed} value of the Higgs and the top-quark masses, set respectively to $m_h =
125$\,GeV and $m_t=173$\,GeV, and that these should not be changed
(unless the user would like to calculate the leading order only).
For the generation of full-fledged Monte-Carlo events an interface to {\tt Pythia 8}~\cite{Sjostrand:2004ef,Sjostrand:2014zea} and {\tt Herwig 7}~\cite{Marchesini:1987cf,Bellm:2015jjp} is provided that is identical to the one in {\tt Powheg-Box-V2/ggHH}.

\section{Results}
\label{sec:results}

Our results were produced for a centre-of-mass energy of 
$\sqrt{s}=13$\,TeV 
using the PDF4LHC15{\tt\_}nlo{\tt\_}30{\tt\_}pdfas~\cite{Butterworth:2015oua}
parton distribution functions interfaced to our code via
LHAPDF~\cite{Buckley:2014ana}, along with the corresponding value for
$\alpha_s$.  The masses of the Higgs boson and the top quark have been fixed,
as in the virtual amplitude, to $m_h=125$\,GeV, $m_t=173$\,GeV and their widths
have been set to zero.
Jets are clustered with the anti-$k_T$ algorithm~\cite{Cacciari:2008gp} as
implemented in the FastJet package~\cite{Cacciari:2005hq,
Cacciari:2011ma}, with jet radius $R=0.4$ and a minimum transverse momentum
$p_{T,\mathrm{min}}^{\rm{jet}}=20$\,GeV. We set the renormalisation and factorisation
scales to $\mu_R=\mu_F=m_{hh}/2$.

\subsection{Total cross sections}

We consider three benchmark points, given in Table~\ref{tab:benchmarks}.
Each of the benchmark points belongs to a characteristic $\mhh$ shape cluster at NLO, as derived in
Ref.~\cite{Capozi:2019xsi}.\footnote{Here, we have updated a subset of the benchmark points identified in  Ref.~\cite{Capozi:2019xsi}, performing the same clustering into seven characteristic shapes of the $\mhh$ distribution with the help of unsupervised machine learning applied to HEFT results. However, we extracted central benchmarks under slightly tighter constraints to reflect more recent experimental measurements~\cite{CMS:2020gsy,ATLAS:2021vrm}, i.e. we used $0.83 \leq \ct \leq 1.17$ for all benchmarks, and additionally $|\ctt| < 0.05$ in the case of benchmark 1.}
The values for $\cgg$ have been modified in order to fulfil the SMEFT relation $\cg=2\cgg$.
Here we consider three out of the seven shape types:
\begin{itemize}
\item benchmark  1: enhanced low $\mhh$ region,
\item benchmark  3: enhanced low $\mhh$ and second local maximum above $\mhh \simeq 2m_t$,
\item benchmark  6: close-by double peaks or shoulder left.
\end{itemize}

\begin{table}[htb]
  \begin{center}
    \begin{footnotesize}
\begin{tabular}{ |c|c|c|c|c|c||c|c|c|c|c| }
\hline
\begin{tabular}{c}
benchmark \\
($^\ast=$ modified)
\end{tabular} & $c_{hhh}$ & $c_t$ & $ c_{tt} $ & $c_{ggh}$ & $c_{gghh}$ & $C_{H,\textrm{kin}}$ & $C_{H}$ & $C_{uH}$ & $C_{HG}$ & $\Lambda$\\
\hline
SM & $1$ & $1$ & $0$ & $0$ & $0$ & $0$ & $0$ & $0$ & $0$ & $1\;$TeV\\
\hline
$1^\ast$ & $5.105$ & $1.1$ & $0$ & $0$ & $0$ & $4.95$ & $-6.81$ & $3.28$ & $0$ & $1\;$TeV\\
\hline
$3^\ast$ & $2.21$ & $1.05$ & $ -\frac{1}{3} $ & $0.5$ & $0.25^\ast$ & $13.5$ & $2.64$ & $12.6$ & $0.0387$ & $1\;$TeV\\
\hline
$6^\ast$ & $-0.684$ & $0.9$ & $ -\frac{1}{6} $ & $0.5$ & $0.25$ & $0.561$ & $3.80$ & $2.20$ & $0.0387$ & $1\;$TeV\\
\hline
\end{tabular}
\end{footnotesize}
\end{center}
\caption{Benchmark points used for the total cross sections and the distributions of the invariant mass of the Higgs-boson pair,
cf.~Table~\ref{tab:corrected_table3} and Figs.~\ref{fig:bp1distributions}--\ref{fig:bp6distributions}. The value of $C_{HG}$ is determined using $\alpha_s(m_Z)=0.118$.
\label{tab:benchmarks}}
\end{table}

In Table \ref{tab:correct_table3} we list total cross sections at 13\,TeV for these three benchmark points for
option (b) with $\Lambda=1$\,TeV,
as well as the ratio to the SM cross section.
The result for the total cross section with option (a) and the result for HEFT are also shown.
\begin{table}[htb]
\begin{center}
\begin{tabular}{| c | c | c |c|c|c|}
\hline
&&&&&\\
  benchmark  &$\sigma_{\rm{NLO}}$[fb]  &K-factor & ratio to  SM & $\sigma_{\rm{NLO}}$[fb] & $\sigma_{\rm{NLO}}$[fb] \\
  & option (b) & option (b) & option (b) & option (a) & HEFT\\
  &&&&&\\
  \hline
  \hline
SM & 27.94$^{+13.7\%}_{-12.8\%}$  & 1.67 & 1 & - &-\\
  \hline
  \hline
  \multicolumn{6}{|c|}{$\Lambda=1$\,TeV}\\
  \hline
  \hline
1$^\star$ & 71.95$^{+20.1\%}_{-15.7\%}$  & 2.06 & 2.58 & -57.64 & 91.62\\
  \hline
3$^\star$ & 68.69$^{+9.4\%}_{-9.5\%}$ & 1.80 & 2.46 & 30.15 & 70.20\\
  \hline 
6$^\star$ & 70.18$^{+18.8\%}_{-15.5\%}$ & 1.83 & 2.51 & 50.82 & 87.9\\
   \hline
   \hline
 \multicolumn{6}{|c|}{$\Lambda=2$\,TeV}\\
   \hline
   \hline
1$^\star$& 14.53$^{+12.6\%}_{-12.2\%}$  &1.62  & 0.52 & 6.44 & -\\
  \hline
  \hline
\end{tabular}
\end{center}
\caption{Total cross sections for Higgs-boson pair production at full
  NLO QCD for three benchmark points and truncation option (b) for $\Lambda=1$\,TeV.
  The total cross sections for truncation option (a) are also given, in order to highlight the difference to the linearised case, as well as the values for HEFT. One can clearly see that truncation option (a) with $\Lambda=1$\,TeV is not a valid option for benchmark point 1$^\star$, leading to an unphysical cross section. 
  The uncertainties are scale uncertainties based on 3-point scale variations.
  The values are the corrected values after the update of the code as described in Section~\ref{sec:erratum}.
\label{tab:correct_table3}}
\end{table}

The squared modulus of the amplitude can be parametrised in terms of all possible
coupling combinations appearing at fixed order, where the $a_i$ have been defined as the coefficients of
coupling combinations in HEFT at NLO QCD~\cite{Buchalla:2018yce},
\begin{align*}
|{\cal M}_{\textrm{BSM}}|^2 =& \,a_1\cdot c_t^4+a_2\cdot c_{tt}^2+a_3\cdot c_t^2 c_{hhh}^2 + a_4\cdot c_{ggh}^2 c_{hhh}^2 + a_5\cdot c_{gghh}^2 + a_6\cdot c_{tt}c_t^2 + a_7\cdot c_t^3 c_{hhh}  \nonumber\\[-2pt]
+&\,a_8\cdot c_{tt} c_t c_{hhh} + a_9\cdot c_{tt} c_{ggh} c_{hhh} + a_{10}\cdot c_{tt} c_{gghh} +a_{11}\cdot c_t^2 c_{ggh} c_{hhh} + a_{12}\cdot c_t^2 c_{gghh}  \nonumber\\[-2pt]
+&\,a_{13}\cdot c_{t} c_{hhh}^2 c_{ggh} + a_{14}\cdot c_{t} c_{hhh} c_{gghh} + a_{15}\cdot c_{ggh} c_{hhh}c_{gghh} +a_{16}\cdot c_t^3 c_{ggh}  \nonumber\\[-2pt]
+&\,a_{17}\cdot c_t c_{tt}c_{ggh} + a_{18}\cdot c_{t} c_{ggh}^2 c_{hhh} + a_{19}\cdot c_{t} c_{ggh} c_{gghh} + a_{20}\cdot c_{t}^2 c_{ggh}^2  \nonumber\\[-2pt]
+&\,a_{21}\cdot c_{tt} c_{ggh}^2 + a_{22}\cdot c_{ggh}^3 c_{hhh} + a_{23}\cdot c_{ggh}^2 c_{gghh}\;.\label{eq:squared_amp}
\end{align*}
Truncation options (a) and (c) of eq.~\eqref{eq:truncation} are expansions at
cross-section level, while (d) is the direct translation from HEFT to SMEFT.
Therefore, for those cases the application of the translation of
Table~\ref{tab:translation}, including all terms at the desired order in
$1/\Lambda$, is sufficient.  For the truncation option (b), there are
combinations which cannot be reconstructed from HEFT and therefore they have
been calculated explicitly and implemented in analytic form.

\begin{figure}[h]
\begin{center}
\includegraphics[width=.47\textwidth,page=1]{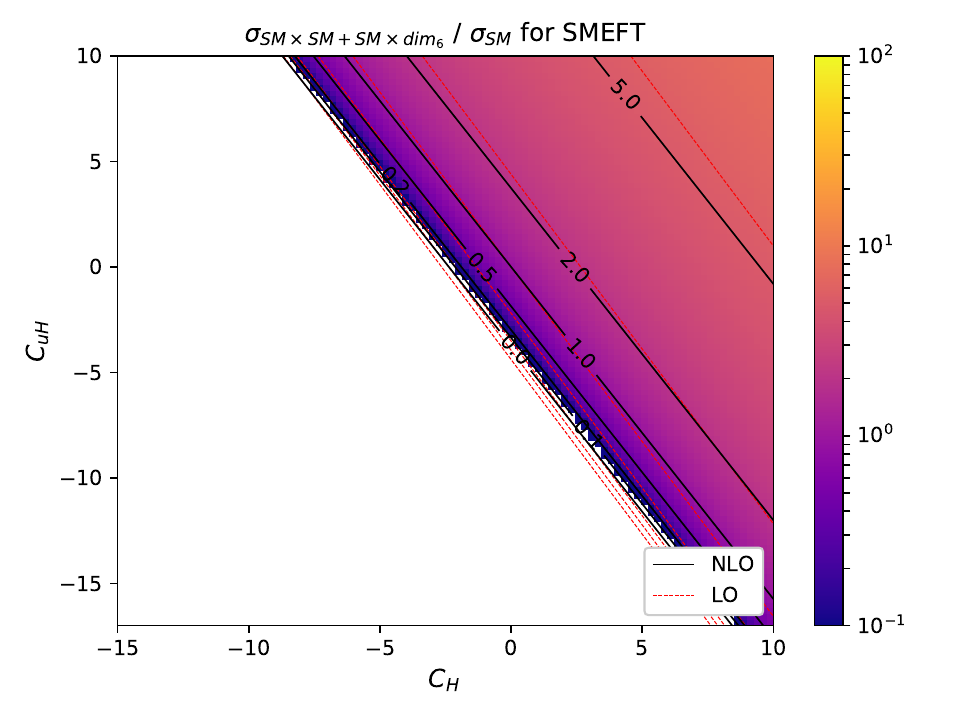}%
\includegraphics[width=.47\textwidth,page=1]{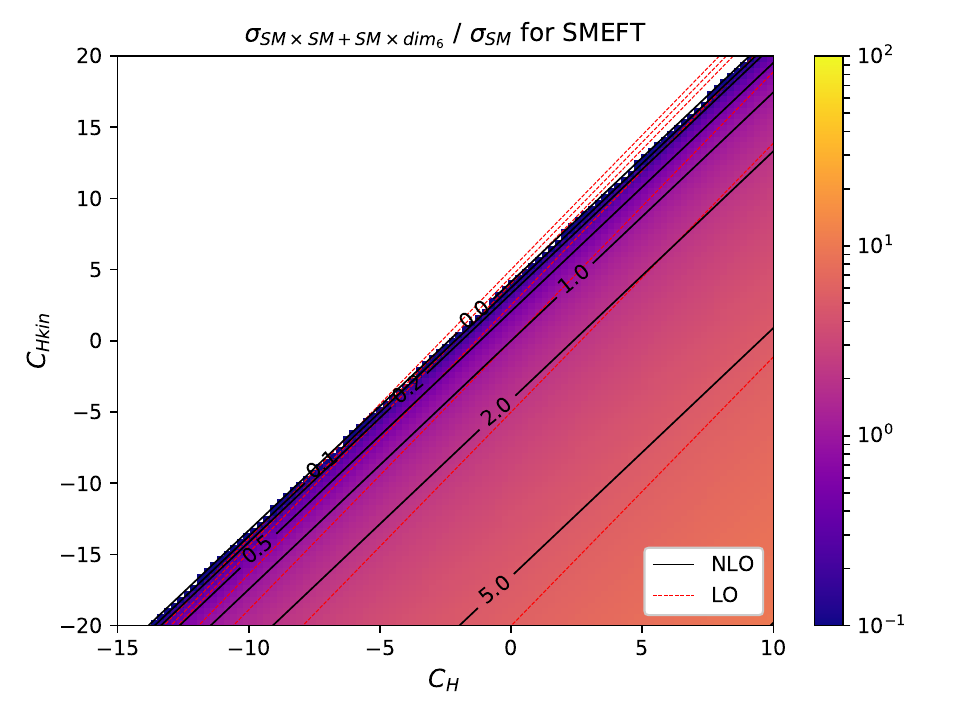}%
\\
\includegraphics[width=.47\textwidth,page=2]{figures/plot_CH_CuH.pdf}%
\includegraphics[width=.47\textwidth,page=2]{figures/plot_CH_CHkin.pdf}%
\\
\includegraphics[width=.47\textwidth,page=4]{figures/plot_CH_CuH.pdf}%
\includegraphics[width=.47\textwidth,page=4]{figures/plot_CH_CHkin.pdf}%
    \caption{\label{heatmaps} Heat maps showing
    the dependence of the cross section on the couplings $C_H$, $C_{uH}$ (left)
    and $C_H$, $C_{H,\textrm{kin}}$ (right) with $\Lambda=1$\,TeV for different truncation
    options. Top: option (a) (linear dim-6), middle: option (b) (quadratic dim-6),
    bottom: option (d) (no linearisation in $1/\Lambda$).
    The white areas denote regions in parameter space where the corresponding cross
    section would be negative.}
\end{center}
\end{figure}

In Fig.~\ref{heatmaps} we show heat maps illustrating the dependence of the
cross section on the couplings $C_H$, $C_{uH}$ (left) and $C_H$, $C_{H,\textrm{kin}}$
(right) with $\Lambda=1$\,TeV for the truncation options (a) (linear dim-6,
upper plots), (b) (quadratic dim-6, middle plots) and (d) (no linearisation, lower plots).  We observe that the
results for the total cross sections (normalised to the SM case) are
substantially different between these  options.
The purely linear dim-6 contributions lead to negative cross sections over
large parts of the parameter space, as manifested by the white areas in the
top row of plots.  This feature is not present at all when we consider the
quadratic dim-6 truncation option.
Furthermore, in the linear dim-6 case, we find completely flat
directions in the considered parameter range for a combined variation of the
respective Wilson coefficients. In the quadratic dim-6 case, option (b), iso-contours  have an elliptic shape due to the
quadratic terms in the cross section, while for option (d) the elliptic iso-contours are distorted due to higher terms in the polynomials in the Wilson coefficients.


\subsection{Distribution of the Higgs-pair invariant mass $m_{hh}$: truncation effects}

We now consider differential results and show the effect of the different
truncation options on the distribution of the invariant mass $\mhh$ of the
Higgs-boson pair.  We present results for the three benchmark points given in
Table~\ref{tab:benchmarks}.
We also show scale uncertainty bands for option (b) and the SM case, resulting from 3-point scale variations around the central scale $\mu_R=\mu_F=c \cdot m_{hh}/2$, with $c \in \lbrace \frac{1}{2}, 1, 2 \rbrace$. For the SM case as well as for benchmark point 1 we have verified that 7-point scale variations lie within the 3-point scale uncertainty band.

\begin{figure}[h]
\includegraphics[width=18pc,page=1]{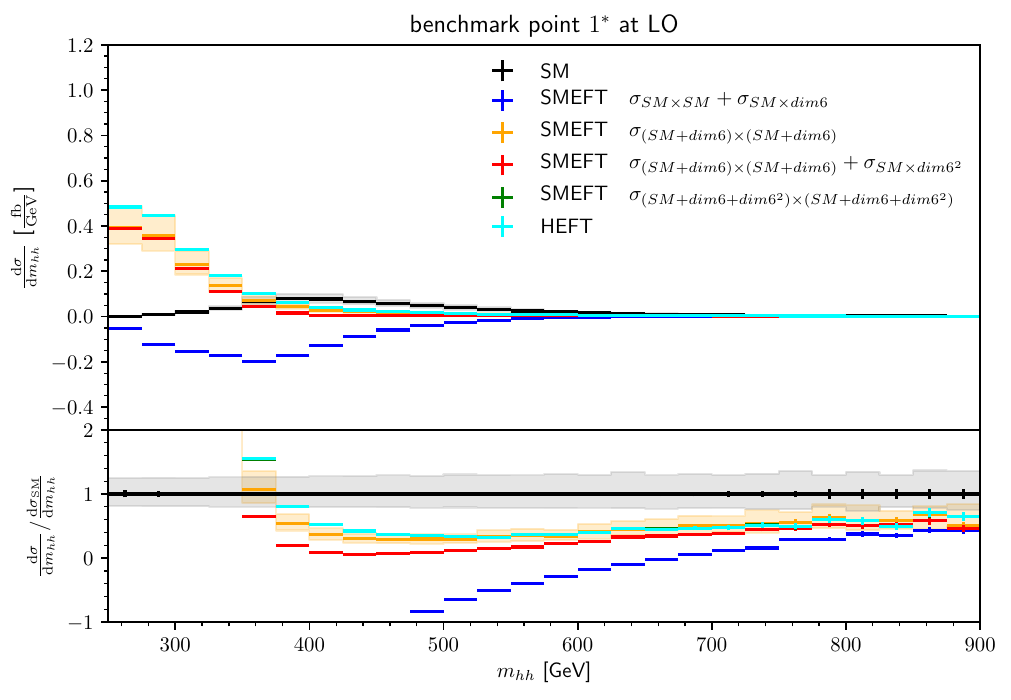}\hspace{2pc}%
\includegraphics[width=18pc,page=1]{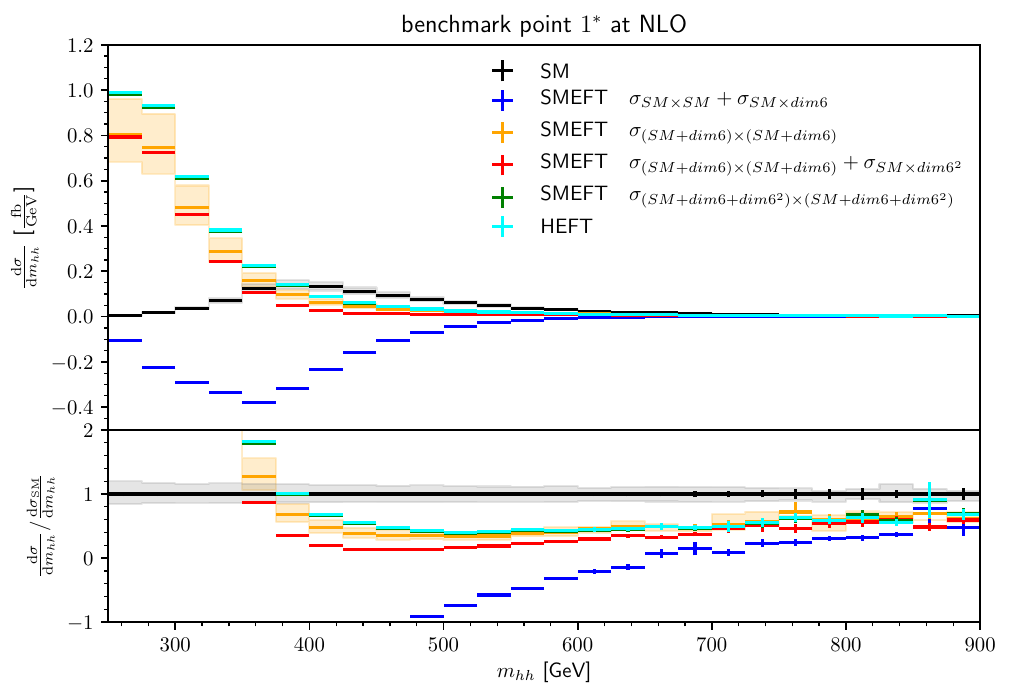}\hspace{2pc}%
\\
\includegraphics[width=18pc,page=1]{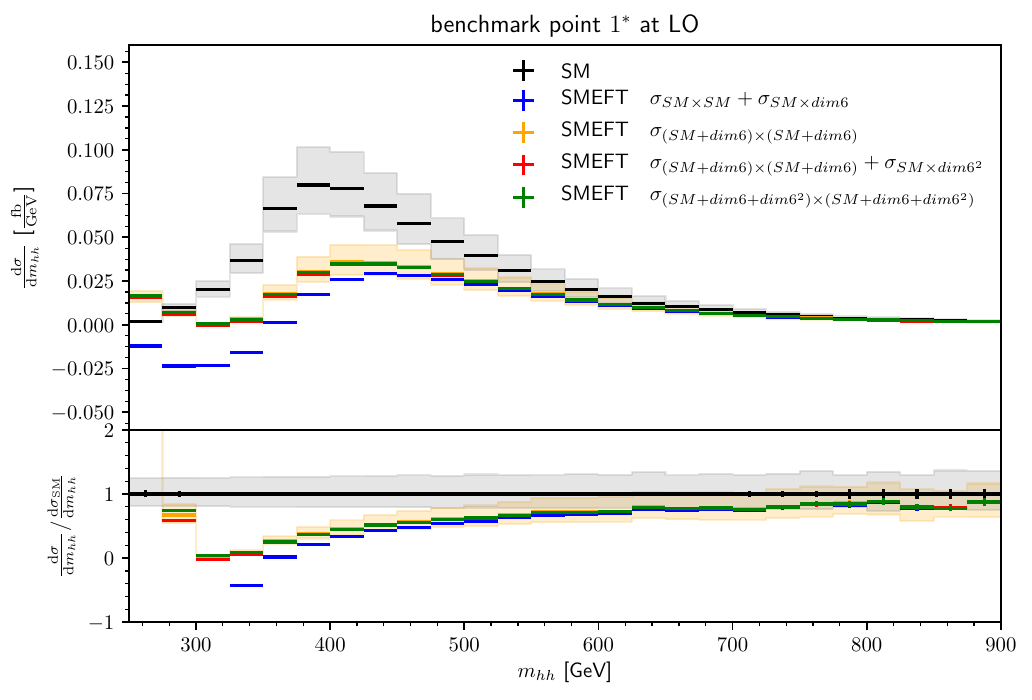}\hspace{2pc}%
\includegraphics[width=18pc,page=1]{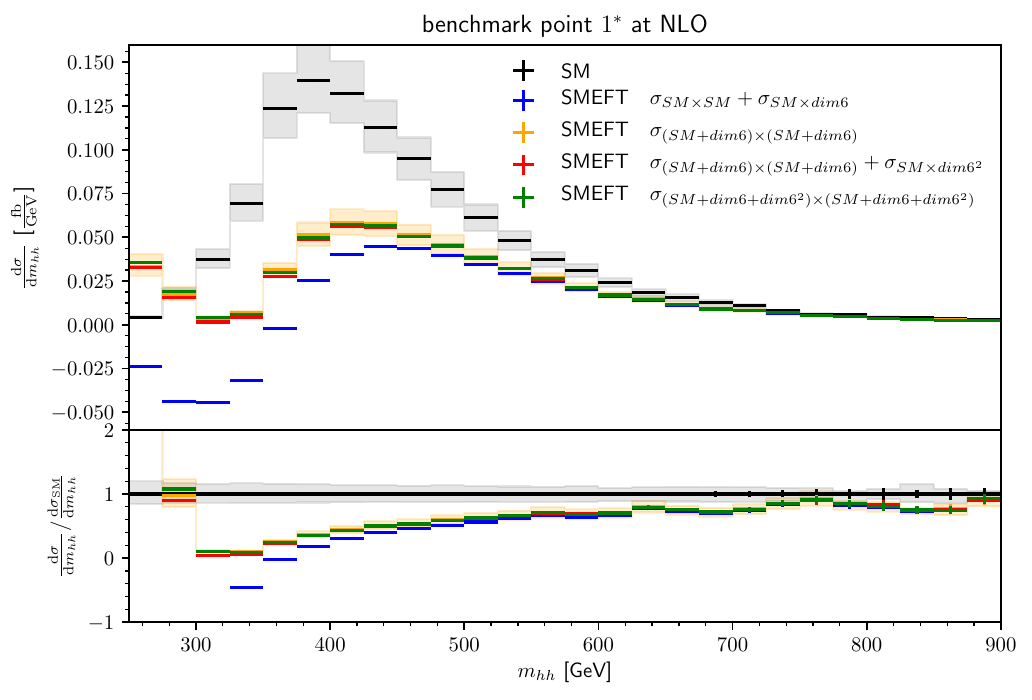}\hspace{2pc}%
\\
\includegraphics[width=18pc,page=1]{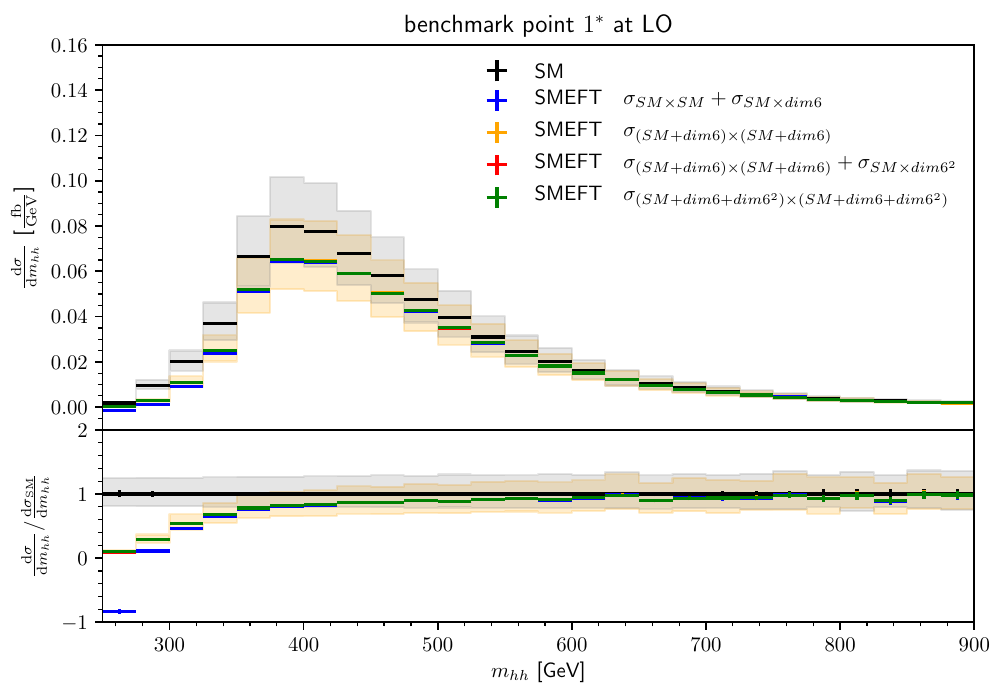}\hspace{2pc}%
\includegraphics[width=18pc,page=1]{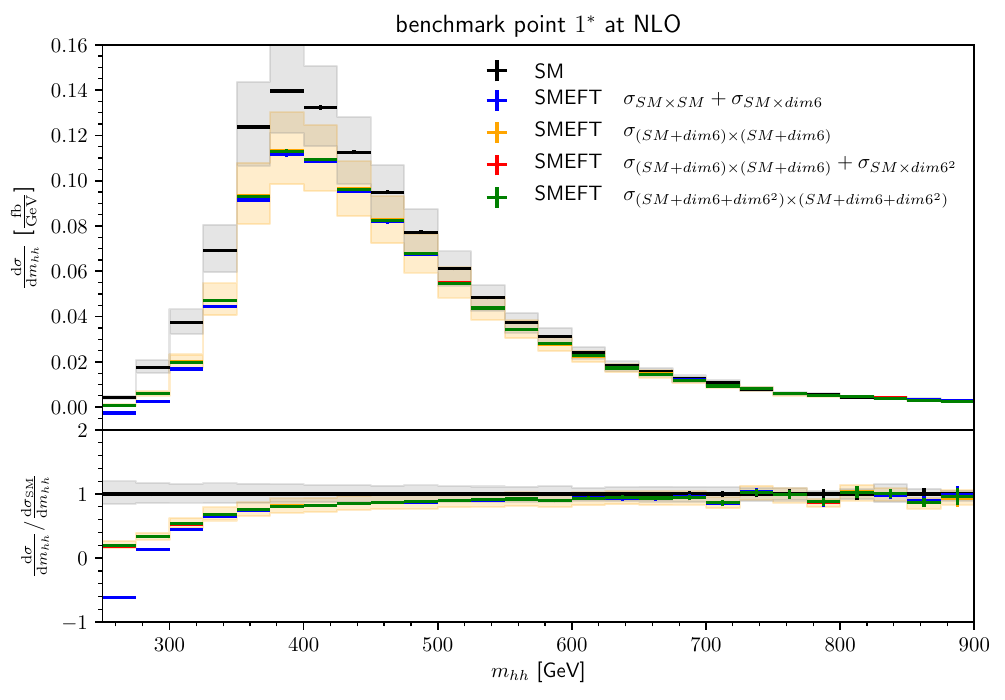}\hspace{2pc}%
   \caption{\label{fig:bp1distributions} Differential cross sections for the invariant mass $\mhh$ of the
   Higgs-boson pair for benchmark point 1 of Table~\ref{tab:benchmarks}. Top row: $\Lambda=1$\,TeV, middle row:
   $\Lambda=2$\,TeV, bottom row:
   $\Lambda=4$\,TeV.  Left: LO, right: NLO. (All the subfigures show the corrected distributions after the update of the code, LO was not affected.)}
\end{figure}

\begin{figure}[h]
\includegraphics[width=18pc,page=1]{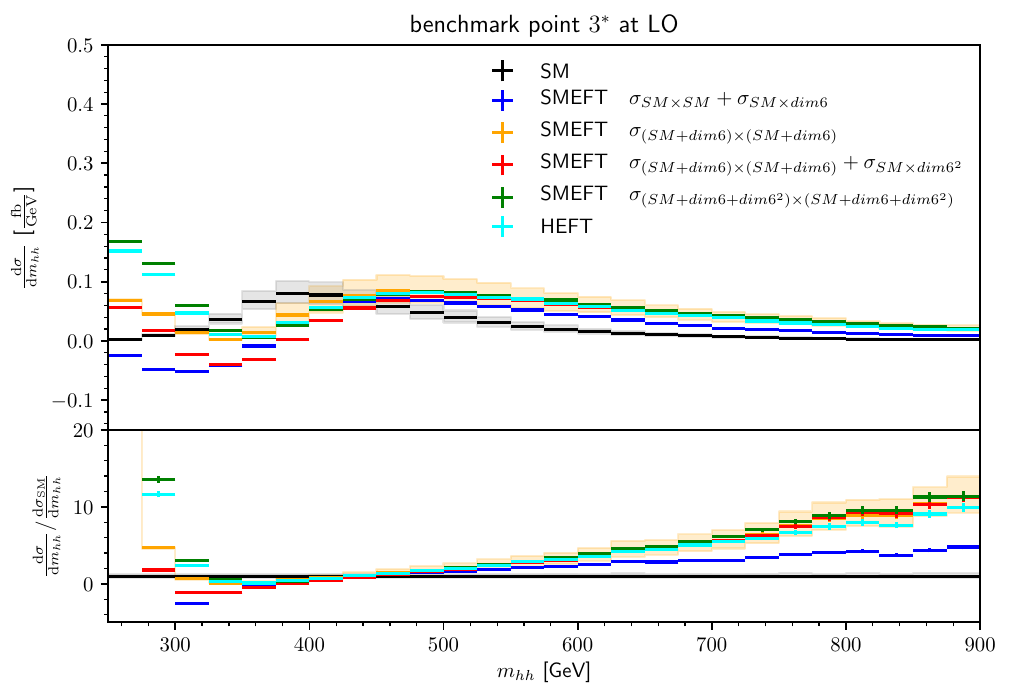}\hspace{2pc}%
\includegraphics[width=18pc,page=1]{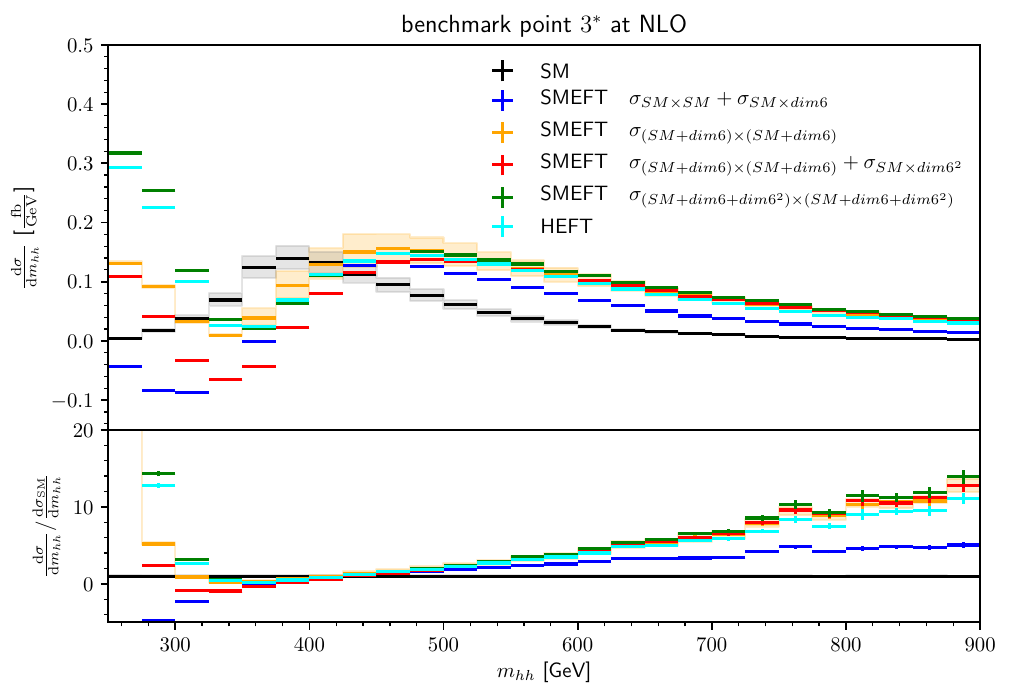}\hspace{2pc}%
\\
\includegraphics[width=18pc,page=1]{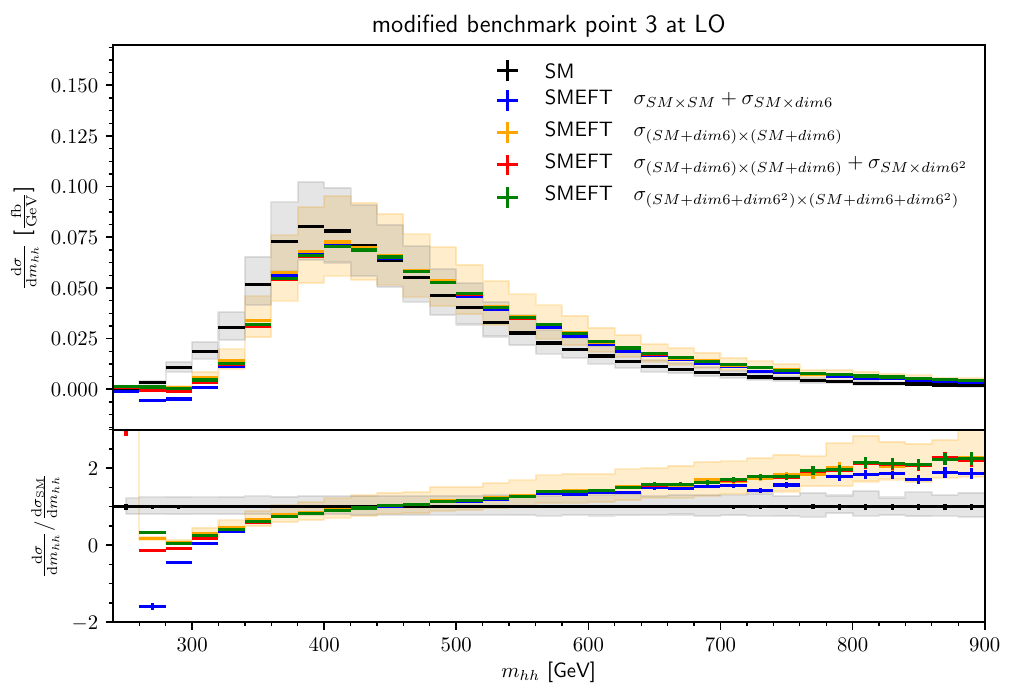}\hspace{2pc}%
\includegraphics[width=18pc,page=1]{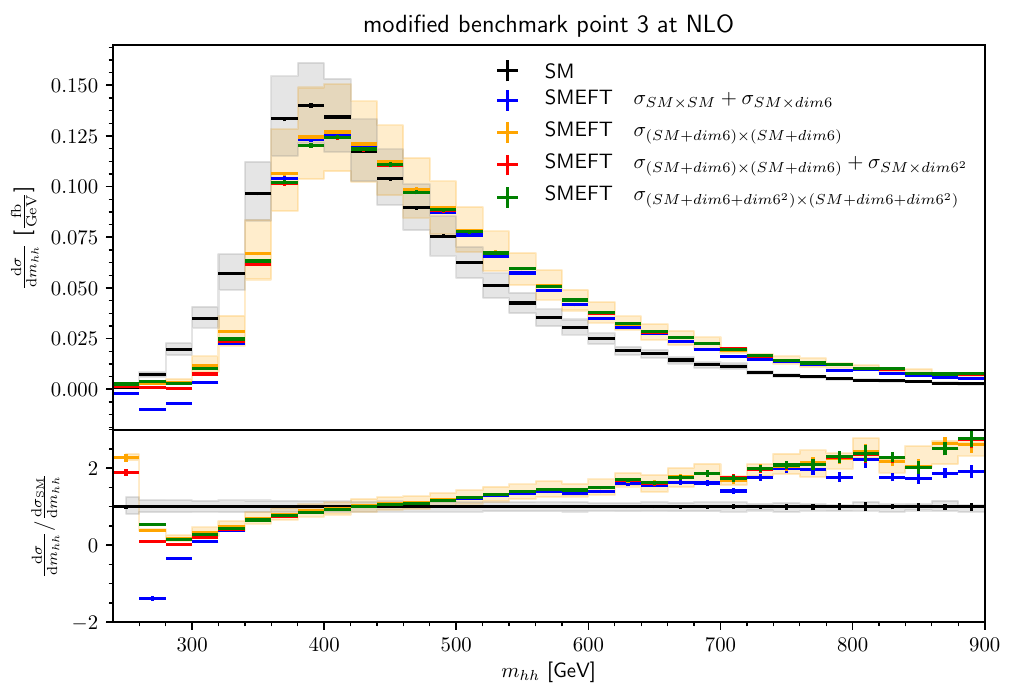}\hspace{2pc}%
\\
\includegraphics[width=18pc,page=1]{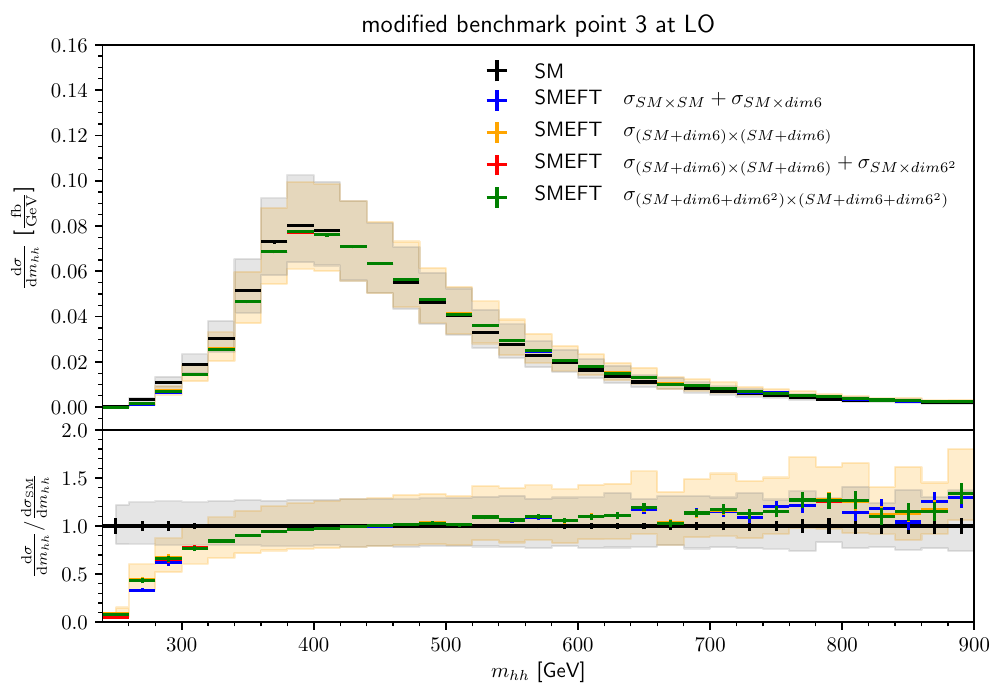}\hspace{2pc}%
\includegraphics[width=18pc,page=1]{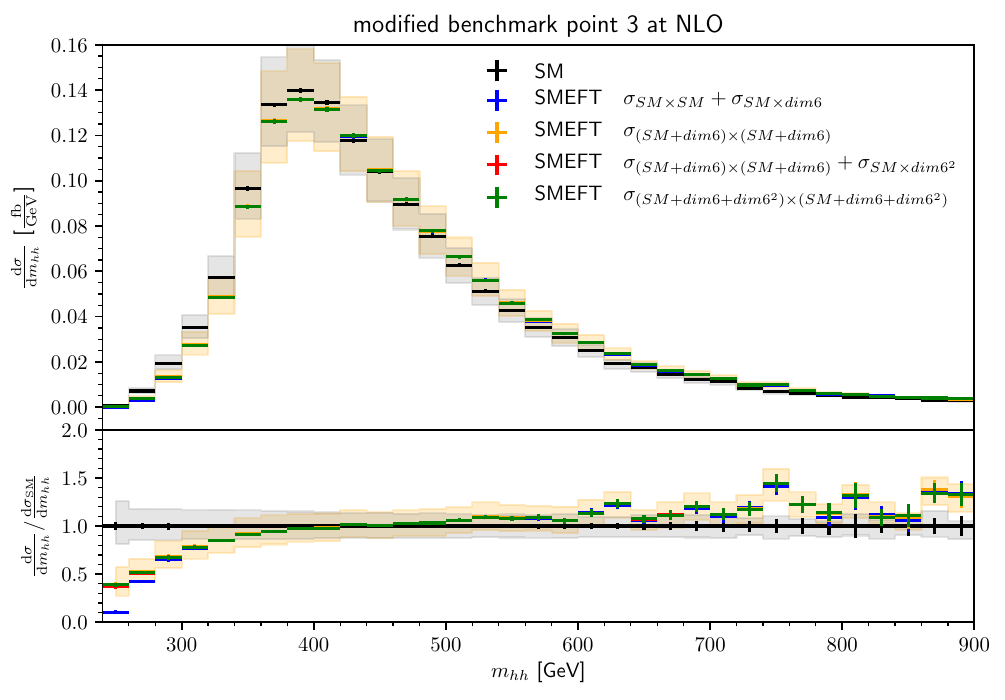}\hspace{2pc}%
   \caption{\label{fig:bp3distributions} Differential cross sections for the invariant mass $\mhh$ of the
   Higgs-boson pair for benchmark point 3 of Table~\ref{tab:benchmarks}. Top row: $\Lambda=1$\,TeV, middle row:
   $\Lambda=2$\,TeV, Bottom row:
   $\Lambda=4$\,TeV.  Left: LO, right: NLO. (Plots resulting from the updated code, except for  $\Lambda=2$\,TeV and  $\Lambda=4$\,TeV at NLO, where the differences are at the sub-percent level.)
 }
\end{figure}

\begin{figure}[h]
\includegraphics[width=18pc,page=1]{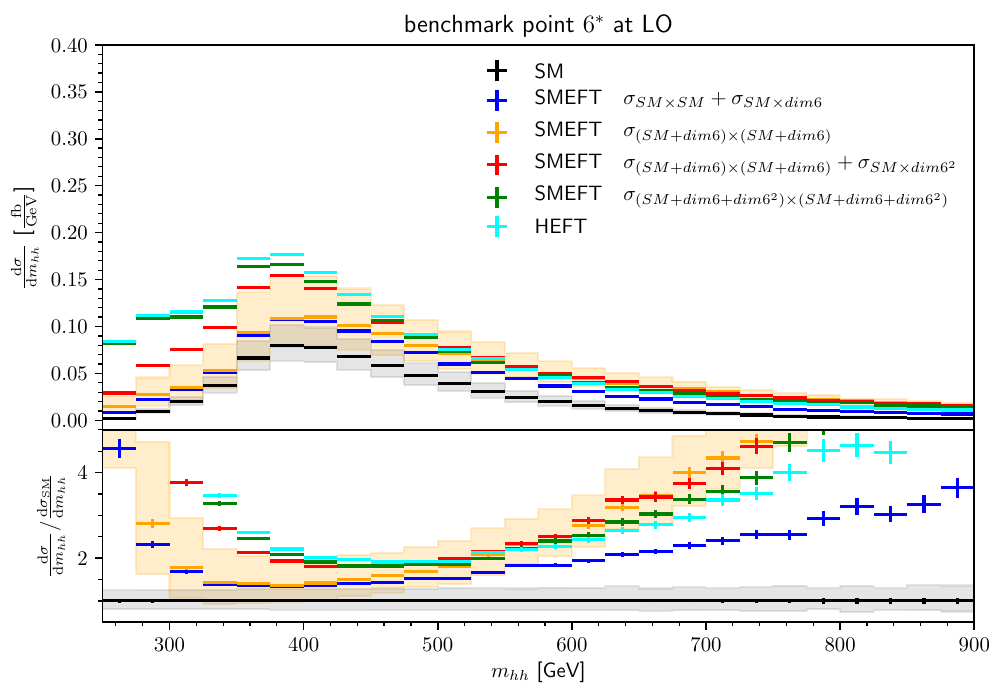}\hspace{2pc}
\includegraphics[width=18pc,page=1]{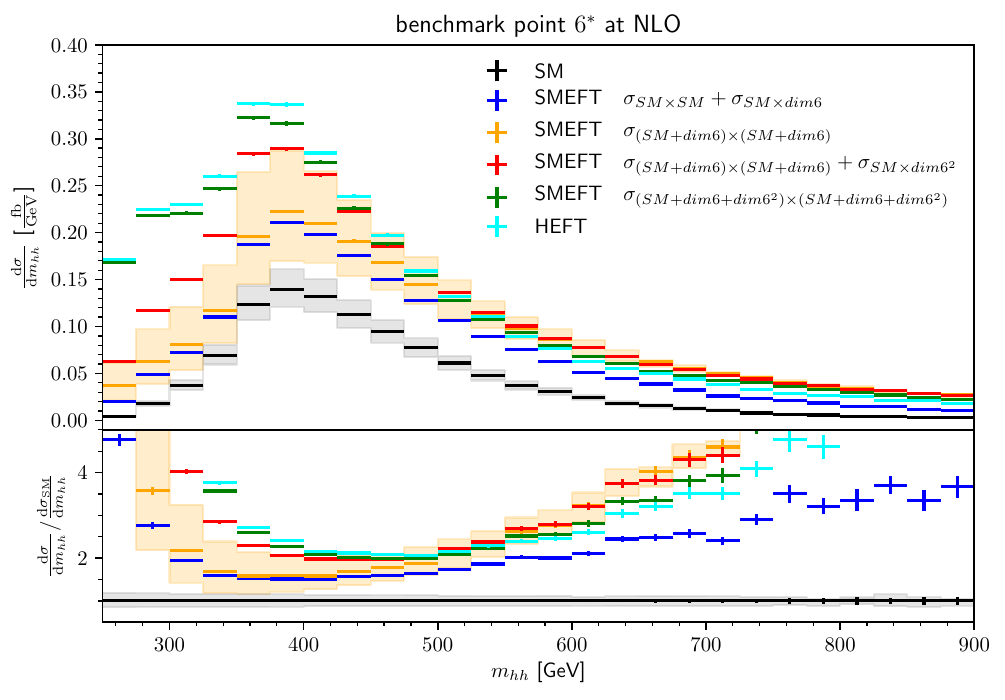}\hspace{2pc}%
\\
\includegraphics[width=18pc,page=1]{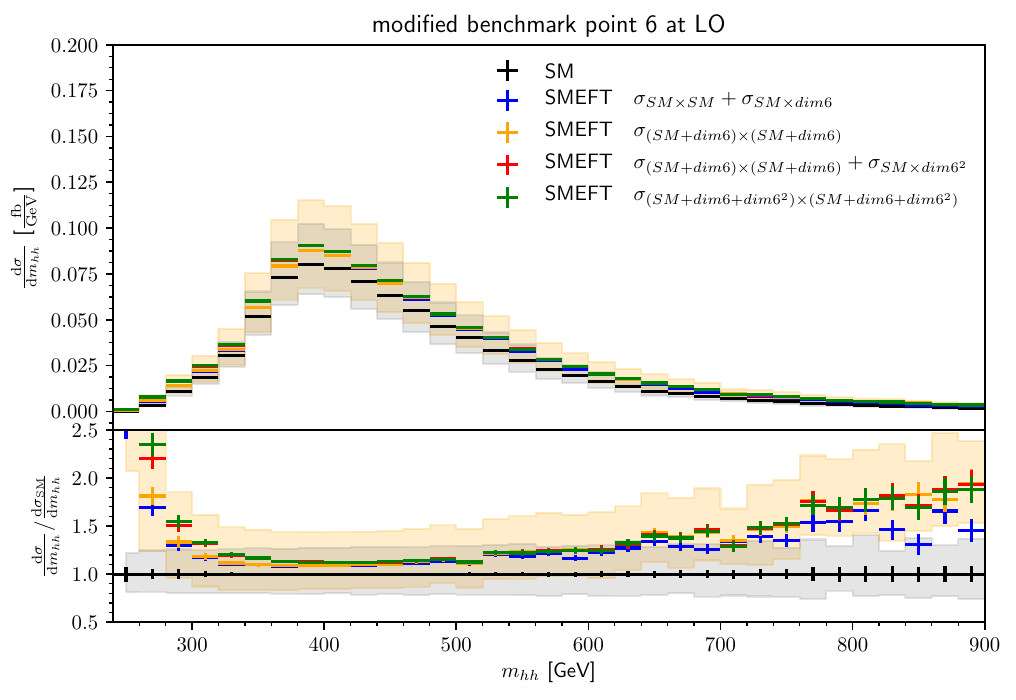}\hspace{2pc}%
\includegraphics[width=18pc,page=1]{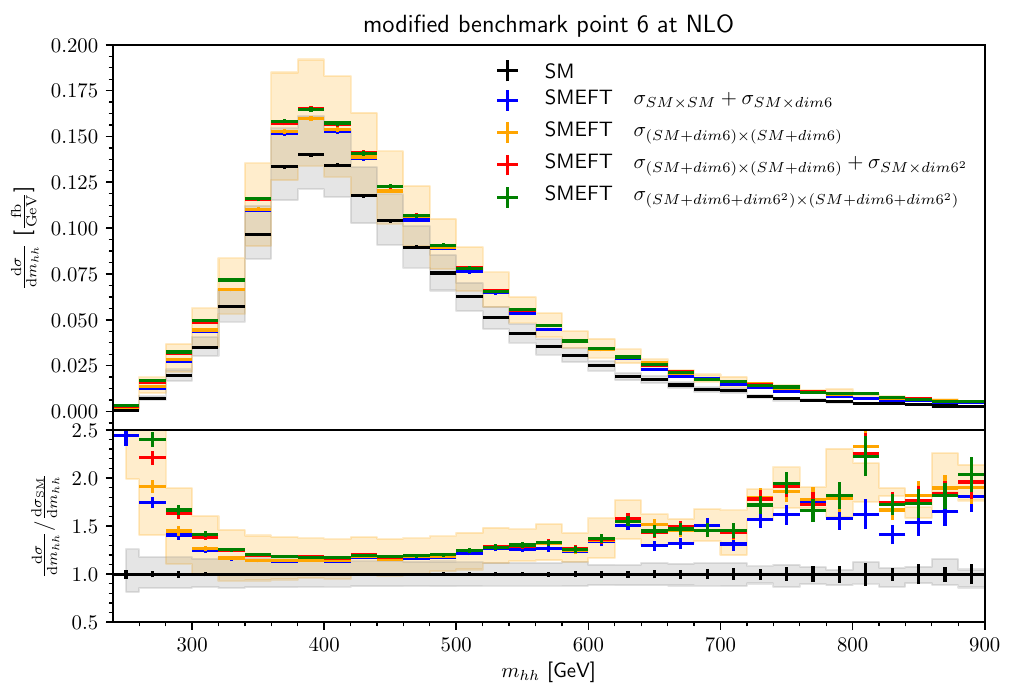}\hspace{2pc}%
\\
\includegraphics[width=18pc,page=1]{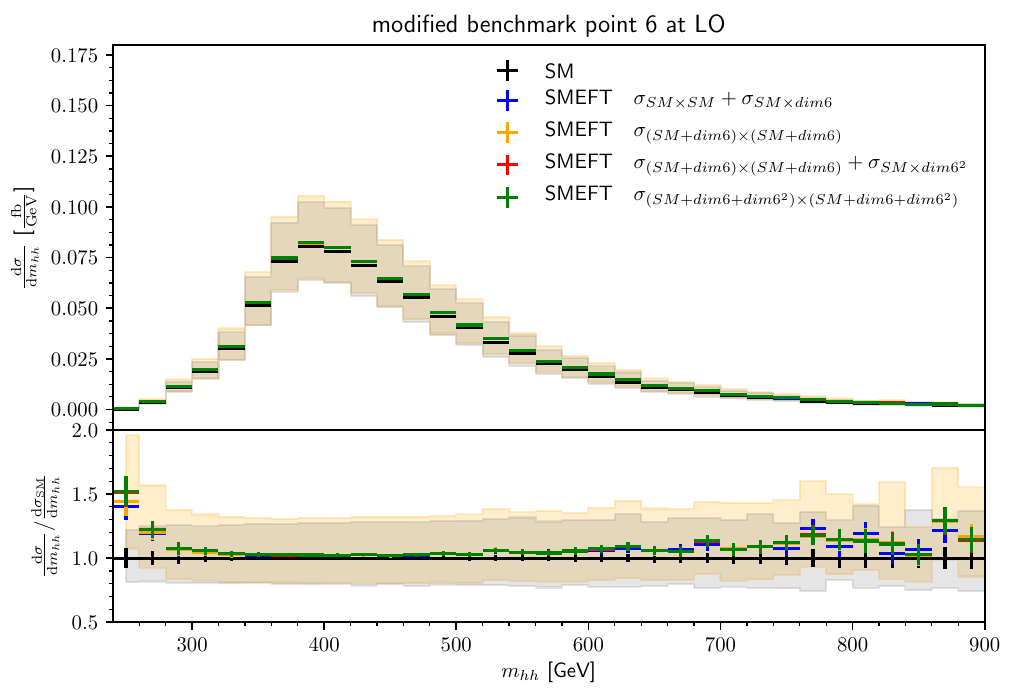}\hspace{2pc}%
\includegraphics[width=18pc,page=1]{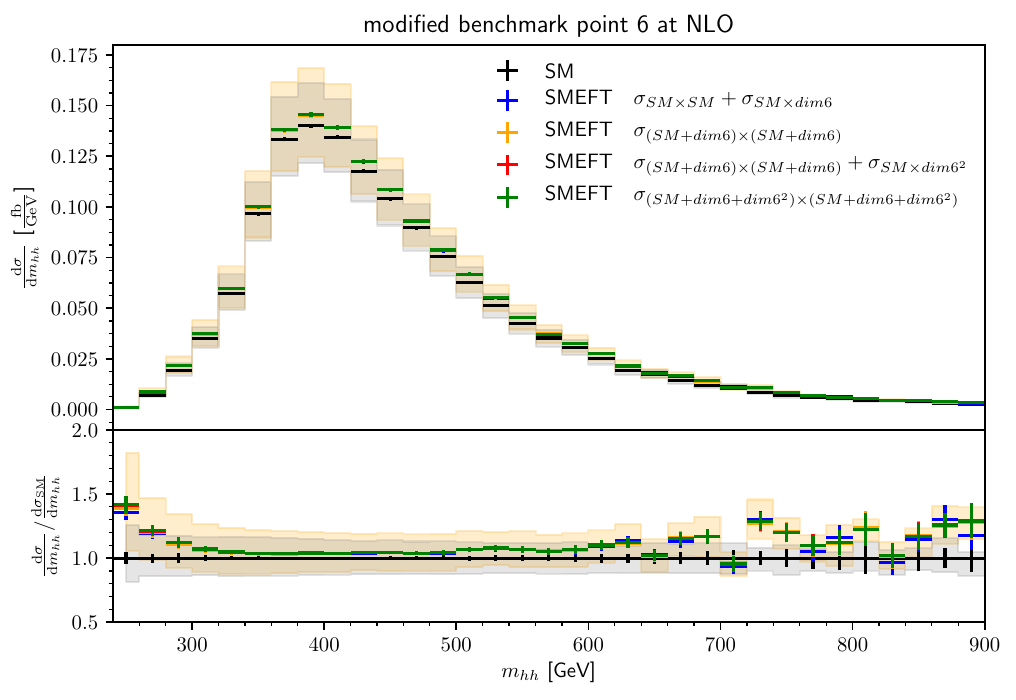}\hspace{2pc}%
   \caption{\label{fig:bp6distributions} Differential cross sections for the invariant mass $\mhh$ of the
   Higgs-boson pair for benchmark point 6 of Table~\ref{tab:benchmarks}. Top row: $\Lambda=1$\,TeV, middle row:
   $\Lambda=2$\,TeV, bottom row:
   $\Lambda=4$\,TeV.  Left: LO, right: NLO.  (Plots resulting from the updated code, except for  $\Lambda=2$\,TeV and  $\Lambda=4$\,TeV at NLO, where the differences are at the sub-percent level.)}
\end{figure}

The SMEFT results for benchmark points 1, 3 and 6 are shown in
Figs.~\ref{fig:bp1distributions},~\ref{fig:bp3distributions} and~\ref{fig:bp6distributions}, respectively, at $\Lambda=1$\,TeV (upper panels), $\Lambda=2$\,TeV (middle panels) and
$\Lambda=4$\,TeV (lower panels), for the different truncation options (a)
through (d), where the left (right) row shows LO (NLO) results.
In the upper panels, we also display the HEFT results for the same
benchmark point, with the translation given numerically in
Table~\ref{tab:benchmarks} for a value of $\Lambda=1$\,TeV.
The dark blue curve
corresponds to the linear dim-6 case, i.e.~truncation option (a).  As mentioned
above, the negative differential cross-section values in the linear dim-6 case
indicate that points in the coupling parameter space which are valid in HEFT
can lead, upon naive translation of the Wilson coefficients, to parameter
points for which the SMEFT expansion is not valid. The orange curve
corresponds to the truncation option (b), where squared dim-6 contributions are
taken into account. This choice, as well as option (c), leads to positive differential cross sections.\footnote{Nevertheless, one can question whether including an incomplete subset of $1/\Lambda^4$ contributions is theoretically well-defined (in particular whether such truncation options leave the theory gauge-invariant~\cite{Trott:2021vqa}). We take the view that these options can at least be useful in determining the magnitude of truncation uncertainties in $gg \to hh$. Furthermore, field redefinition ambiguities are expected to be suppressed in the $gg \to hh$ case as they enter via interference with a loop-induced SM amplitude~\cite{Martin:2021cvs}.}

For benchmark point 1, the characteristic shape in HEFT features an enhanced low-$\mhh$ region.
From Fig.~\ref{fig:bp1distributions} we see that this shape is not preserved in SMEFT as $\Lambda$ increases.
Disregarding the fully linearised option, which is simply not a valid option for this benchmark point, we see that the distribution develops a dip for
$\Lambda=2$\,TeV;
as the other couplings
are almost equal to the SM case, increasing $\Lambda$ translates to decreasing the effective trilinear coupling, which becomes numerically close to the value of maximal destructive interference between box- and triangle-contributions. As the heavy scale is increased further to $\Lambda=4$\,TeV, the distribution approaches the SM shape.
Note that for benchmark point 1, where $\cg$ and $\cgg$ are zero, curves from option (d) (green) and HEFT (cyan) are identical at  $\Lambda=1$\,TeV. 

For benchmark point 3, the differences between the truncation options are very pronounced, see Fig.~\ref{fig:bp3distributions}.
The value of $\chhh$ is close to the value of maximal destructive interference between box- and triangle-type contributions when considering the cross section as a function of $\chhh$ alone, therefore delicate cancellations are likely to take place.
Again, the $\mhh$-shape changes as $\Lambda$ increases from 1\,TeV to 2\,TeV.

Furthermore, we observe that the contribution from the interference of double
dimension-6 operator insertions with the SM, which appeared to be sub-dominant even for $\Lambda=1$\,TeV in the
case of benchmark point 1, modifies the interference pattern considerably  in the case of benchmark point 3,
as can be seen by comparing truncation option (b) (orange)
with option (c) (red), where the latter includes the double operator insertions
interfered with the SM amplitude.
While for $\Lambda=1$\,TeV   the differences  between the truncation options are large, benchmark point 3 shows a faster convergence to the SM shape as $\Lambda$ increases. For $\Lambda=4$\,TeV, the scale uncertainty bands largely overlap with the SM uncertainty bands, except at very low and very high $\mhh$.

For benchmark point 6, the pattern of destructive interference between
different parts of the amplitude (e.g.~box- and triangle-type diagrams) in HEFT
is similar to the one in the SM case. However, in SMEFT (taking the squared
dim-6 level -- option (b) -- as reference), this interference pattern is modified, leading to a
 smaller cross section than in HEFT.
 Furthermore, the characteristic shape (see Fig.~\ref{fig:bp6distributions}) is not preserved for any of the considered $\Lambda$ values:
 in HEFT, the characteristic feature of benchmark 6 is a shoulder left.
 In SMEFT, this shoulder is absent (except for option (d) and $\Lambda=1$\,TeV, which corresponds to HEFT apart from the running of $\alpha_s$).

Looking at the explicit values of the SMEFT coupling parameters in
Table~\ref{tab:benchmarks}, stemming from the naive translation at
$\Lambda=1$\,TeV between HEFT and SMEFT, it becomes clear that the parameters
are too large for the SMEFT expansion  to be valid. Therefore
the large differences seen in the results at $\Lambda=1$\,TeV cannot be regarded as a truncation
uncertainty.  
However, for
$\Lambda=2$\,TeV these values are divided by a factor of 4 and for
$\Lambda=4$\,TeV by a factor of 16, the latter case leading to perfectly valid
SMEFT points for all three benchmarks.
Except for benchmark 1, at $\Lambda=4$\,TeV, the scale uncertainty bands overlap with the SM case for almost all the $\mhh$ range, so it seems that the parameter space where the SMEFT expansion is valid but still clearly distinguishable from the SM case (within NLO uncertainties) is rather small.
This implies that the variety of characteristic $\mhh$-shapes is diminished in SMEFT compared to HEFT.

\section{Conclusions}
\label{sec:conclusions}

We have presented the full NLO QCD corrections to Higgs-boson pair production in gluon fusion within the SMEFT framework. The corresponding matrix elements have been implemented in the {\tt Powheg-Box-V2}\footnote{The code can be downloaded from https://powhegbox.mib.infn.it/ under User-Processes-V2/ggHH\_SMEFT.} in a flexible way, and allow the user to investigate the impact of different truncation options of the series in inverse powers of the new physics scale $\Lambda$.
In particular, we have compared truncations in the inclusion of dimension-6 operators both at amplitude- as well as at cross-section level, and we have studied the effects of double operator insertions into single Feynman diagrams.
We have also compared results from non-linear effective field theory (HEFT) to the different SMEFT truncation options for three benchmark points. 
While some of these truncation options only involve a subset of operators to be included in a consistent counting scheme, truncating the EFT expansion  at the order $1/\Lambda^2$ either at cross-section level, or at amplitude level (where the latter option leads to $1/\Lambda^4$-terms at cross-section  level) are both options which are typically employed in fits to experimental data and which are useful for an estimation of truncation uncertainties.

It is well known that for values of the Wilson coefficients $C_i\,E^2/\Lambda^2\ll 1$, the differences between the considered truncation options should be small for energies typical for LHC processes, i.e.~$E\lesssim 1$\,TeV. However, Higgs-boson pair production in gluon fusion is a process with delicate cancellations between different parts of the amplitude, such that small differences in the treatment of the Wilson coefficients can have a rather large effect, especially in differential distributions.

Furthermore, we have shown that a naive translation between HEFT and SMEFT Wilson coefficients, based on a comparison of terms at Lagrangian level, has to be done with great care (if at all).
The two counting schemes rely on different assumptions, and for the SMEFT expansion to be valid, the values of $C_i\,E^2/\Lambda^2$ should be small. However, starting from a point in the space of anomalous couplings in HEFT that fulfils all the current constraints from measurements, after naive translation one easily ends up at values of $C_i$ where the SMEFT expansion in $C_i\,E^2/\Lambda^2$ is not valid for $E \gtrsim \mhh$
and $\Lambda\simeq 1$\,TeV, the latter being a value often used in the literature for this purpose.

We have illustrated the effects of such translations and of the different truncation options both at total cross-section level as well as for the $\mhh$ distribution, at three benchmark points which are characteristic for a certain $\mhh$-shape in HEFT.
In SMEFT, it is clear that the shape must change for fixed $C_i$ as a function of $\Lambda$, as the SM shape is always approached for large values of $\Lambda$.
At $\Lambda=4$\,TeV, two of the three benchmark points lead to a $\mhh$ distribution that is so close to the SM shape that it is indistinguishable from the SM within the NLO scale uncertainties.

\section{Addendum}
\label{sec:erratum}

\begin{figure}[h]
 \includegraphics[width=18pc,page=1]{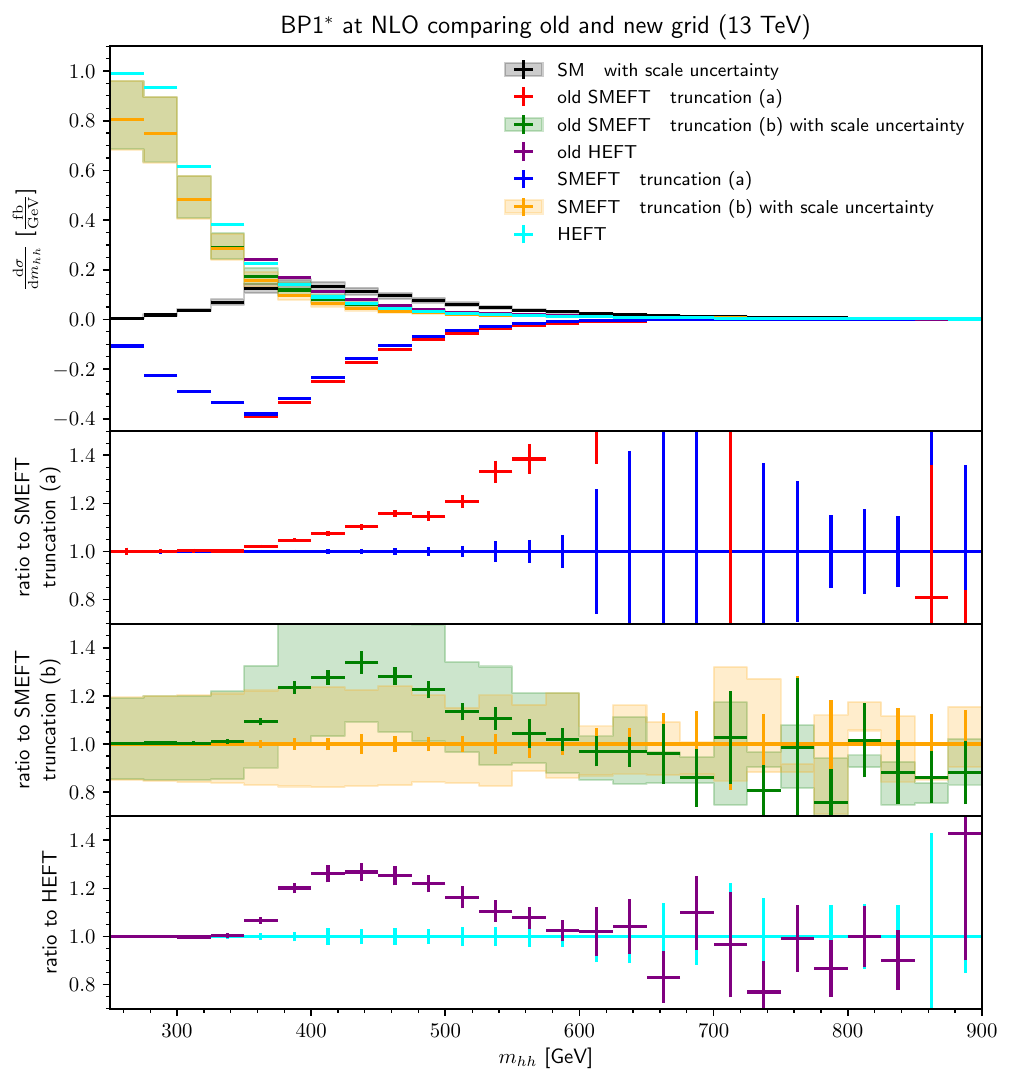}\hspace{2pc}%
\includegraphics[width=18pc,page=1]{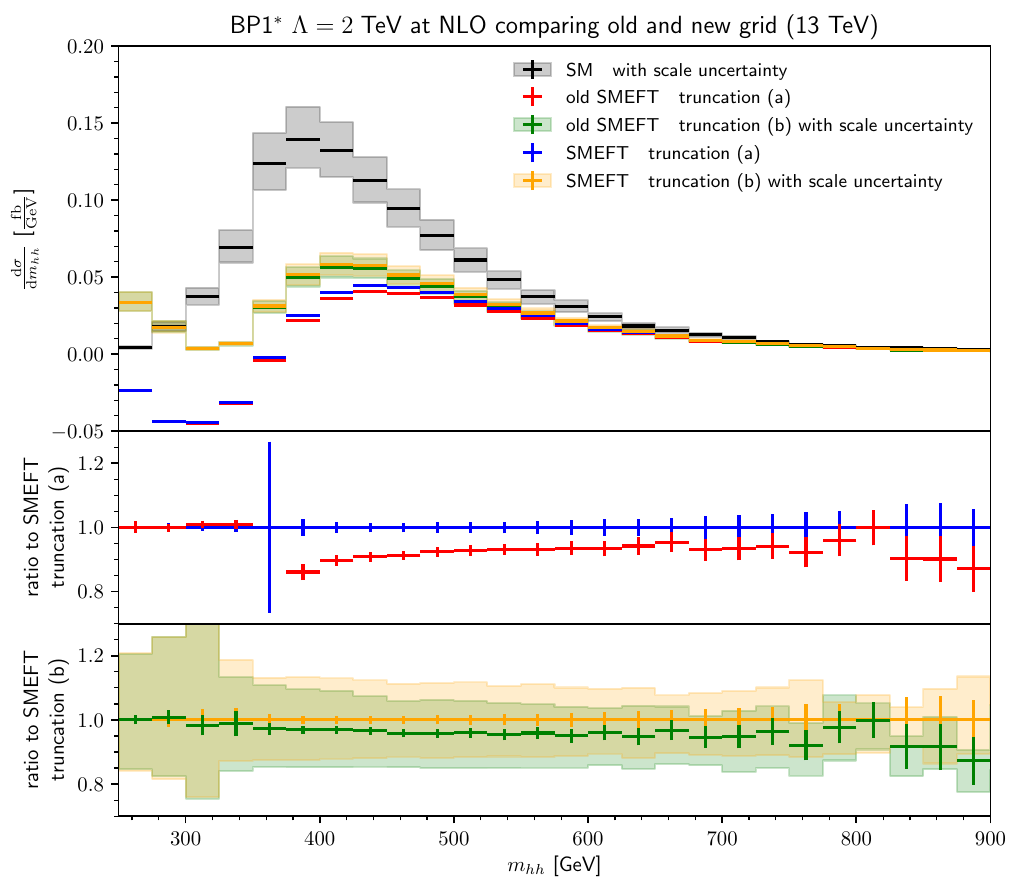}\hspace{2pc}%
  \caption{\label{fig:compare_old_new1} 
     Comparison of old and new results for the cross sections differential in $\mhh$ for benchmark point 1$^\star$,
     with $\Lambda=1$\,TeV (left) and  $\Lambda=2$\,TeV (right), for truncation options (a) and (b). The HEFT distributions for benchmark point 1$^\star$ are also included in the left plot. 
     The lower panels show the truncation options separately and normalised to the corrected result (with 3-point scale variations for option (b)).}
\end{figure}

\begin{figure}[h]
 \includegraphics[width=18pc,page=1]{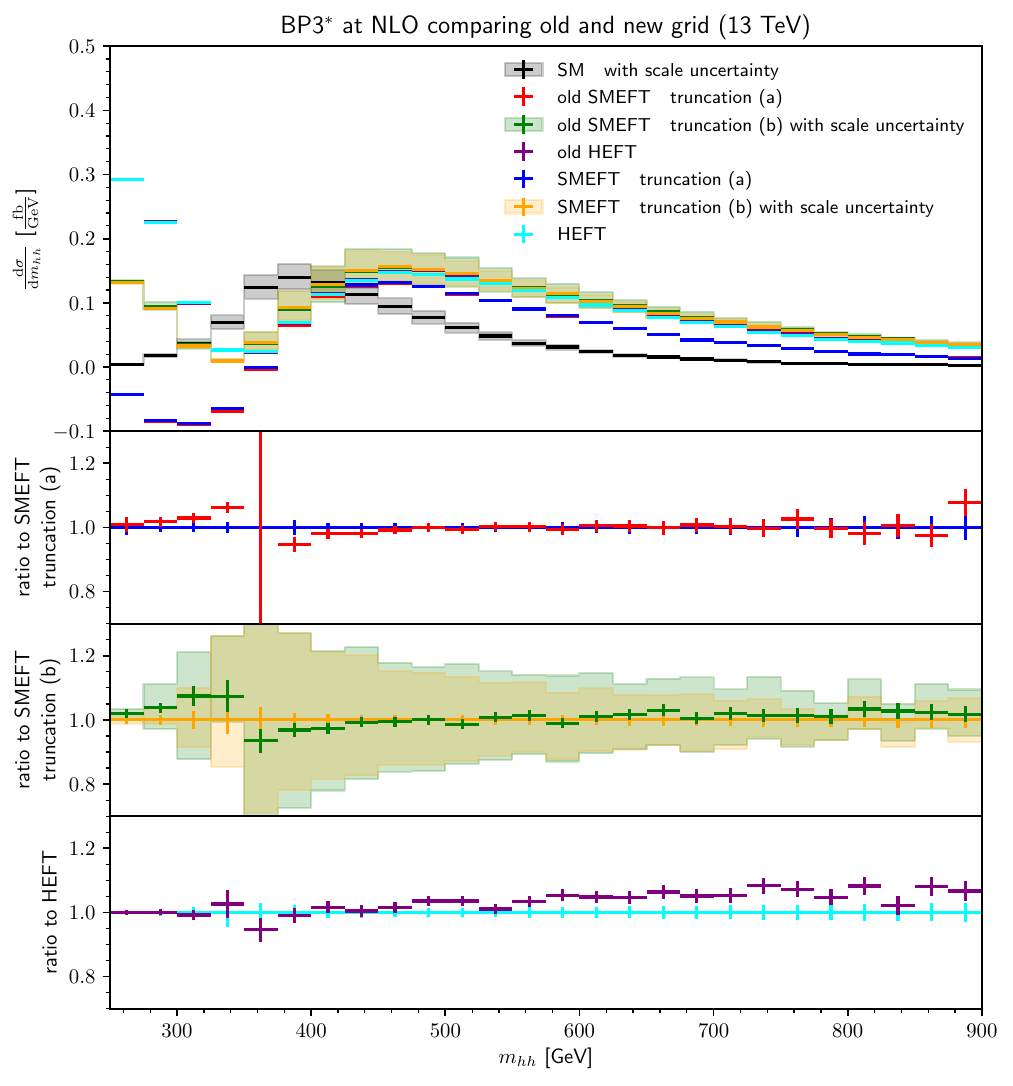}\hspace{2pc}%
\includegraphics[width=18pc,page=1]{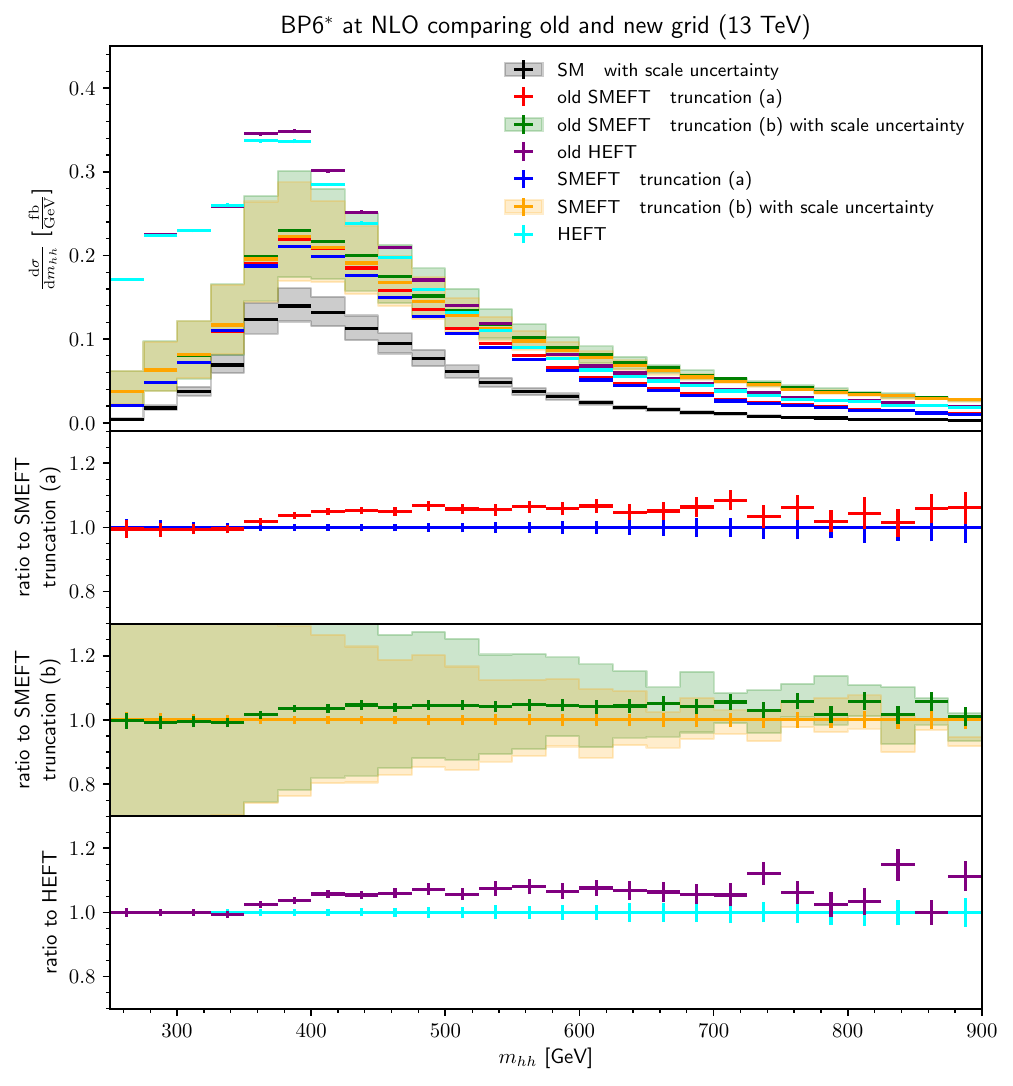}\hspace{2pc}%
   \caption{\label{fig:compare_old_new2} 
     Comparison of old and new results for the cross sections differential in $\mhh$ for benchmark points 3$^\star$ and 6$^\star$,
     with $\Lambda=1$\,TeV and truncation options (a) and (b) and HEFT. 
     The lower panels show the truncation options separately and normalised to the corrected result (with 3-point scale variations for option (b)).}
\end{figure}

The current arXiv version contains the correction of a mistake that has been
reported in the Erratum https://doi.org/10.1007/JHEP10(2023)086:

After comparison with the authors of Ref.~\cite{Bagnaschi:2023rbx}, it turned out
that the two-loop amplitude used in the original version of this
manuscript was
missing a term related to triangle-type two-loop diagrams, affecting the cases
where the ratio between trilinear Higgs coupling $c_{hhh}$ and Yukawa coupling modifier $c_t$ is different from 1 (i.e. the Standard Model (SM) value), 
or when the effective coupling of a $t\bar{t}$ pair to a Higgs pair, $c_{tt}$, is nonzero.
The SM results are unchanged.
Therefore, benchmark points with a value of $c_{hhh}/c_{t}$ or
$c_{tt}$ very different from the SM show the largest difference, which
is up to 35\% for benchmark point 1$^\star$ in the kinematic range
near $m_{hh}=450$\,GeV for truncation option (b), see
Fig.~\ref{fig:compare_old_new1}.
For the other truncation options and for HEFT the qualitative behaviour is similar.
For  benchmark points 3$^\star$ and 6$^\star$ the differences are
below 10\% and therefore within the scale uncertainties, as shown in Fig.~\ref{fig:compare_old_new2}.

The corrected figure for benchmark point 1$^\star$  is shown in the
main text in Fig.~\ref{fig:bp1distributions}.
For benchmark points 3$^\star$ and 6$^\star$ we show the corrected
plots for $\Lambda=1$\,TeV in Figs.~\ref{fig:bp3distributions} and~\ref{fig:bp6distributions}.

\vspace*{3mm}

\section*{Acknowledgements}
We are grateful to the authors of Ref.~\cite{Bagnaschi:2023rbx} for pointing us to the discrepancy with their result.\\
We also would like to thank Stephen Jones and Matthias Kerner for
collaboration related to the $ggHH$@NLO project and Gerhard
Buchalla and  Michael Trott for useful discussions. 
This research was supported by the Deutsche Forschungsgemeinschaft (DFG, German Research Foundation) under grant 396021762 - TRR 257.
LS is supported by the Royal Society under grant number
RP\textbackslash R1\textbackslash 180112 and by Somerville College.

\clearpage
\bibliographystyle{JHEP}
\bibliography{main_gghh_smeft}

\end{document}